# Smart Web Services (SmartWS) – The Future of Services on the Web

Maleshkova, Maria; Philipp, Patrick; Sure-Vetter, York and Studer, Rudi

*The past few years have been marked by an increased use of sensor technologies, abundant availability of mobile devices, and growing popularity of wearables, which enable the direct integration of their data as part of rich client applications. Despite the potential and added value that such aggregate applications bring, the implementations are usually custom solutions for particular use cases and do not support easy integration of further devices. To this end, the vision of the Web of Things (WoT) is to leverage Web standards in order to interconnect all types of devices and real-world objects, and thus to make them a part of the World Wide Web (WWW) and provide overall interoperability. In this context we introduce Smart Web Services (SmartWS) that not only provide remote access to resources and functionalities, by relying on standard communication protocols, but also encapsulate 'intelligence'. Smartness features can include, for instance, context-based adaptation, cognition, inference and rules that implement autonomous decision logic in order to realize services that automatically perform tasks on behalf of the users, without requiring their explicit involvement. In this paper, we present the key characteristics of SmartWS, and introduce a reference implementation framework. Furthermore, we describe a specific use case for implementing SmartWS in the medical domain and specify a maturity model for determining the quality and usability of SmartWS.*

## 1. Introduction

Current developments on the Web are characterized by the wider use of network-enabled devices, such as sensors, mobile phones, and *wearables* that serve as data providers or actuators, in the context of client Web applications [1]. Even though real-life objects can finally participate in Web scenarios, the use of individual and specific interaction mechanisms and data models lead to realizing isolated islands of connected devices or to custom solutions that are not reusable. Devices are increasingly network-enabled but rely on heterogeneous network communication mechanisms, use non-standardized interfaces and introduce new data schemas for each individual type of device [2]. This results in a lot of heterogeneity, in the lack of overall integration and in solutions that cannot easily be extended and reused for different application domains.

We witness similar developments in individual application areas such as in the medical, mobility and energy fields, which face these and even further difficulties. In particular, the situation is aggravated by the growing use of sensors, designated devices for monitoring and recording data, and the digitalization of domain-specific knowledge, such as recordings of trials, guidelines, common procedures, etc. This results in large data volumes [3], which are hard to integrate, process and manage by domain specialists as part of their daily tasks. As a consequence, not only is it difficult to benefit from the available data in order to solve a particular problem or task, it becomes hardly possible to have an overview of all the related information or to keep up with updates.

To this end, the vision of the Web of Things (WoT) [4] is to leverage Web standards in order to interconnect all types of embedded devices (e.g., patient monitors, medical sensors, congestion monitoring devices, traffic-light controls, temperature sensors, smart meters, etc.) and real-world objects, and thus to make them a part of the World Wide Web (WWW) and provide overall interoperability. Therefore, WoT aims to build a future Web of devices that is truly open, flexible, and scalable. We aim to contribute towards achieving this goal by relying on existing and well-known Web standards and paradigms used in the programmable Web (e.g., Uniform Resource Identifiers (URIs), Representational State Transfer (REST) [5], and Hypertext Transfer Protocol (HTTP)) and employing semantic Web technologies (e.g., Resource Description Framework (RDF) [6], and Linked Data (LD) [7]) in order to address the need for

Manuscript published January, 2016. This work was supported in part by the German Research Foundation (DFG) within projects I01, A01, R01, S01 and I04, SFB/TRR 125 "Cognition-Guided Surgery".

Maria Maleshkova (contact author), Patrick Philipp, York Sure-Vetter and Rudi Studer are with the Institute of Applied Informatics and Formal Description Methods (AIFB) and the KSRI - Karlsruhe Service Research Institute, Karlsruhe Institute of Technology (KIT), Germany (e-mail: {maria.maleshkova, patrick.philipp, york.sure-vetter, rudi.studer}@kit.edu).

semantically integrating data coming from a variety of heterogeneous sources and for managing the ever increasing data volumes.

In order to provide a solution for addressing the challenges described above, and to be able to benefit from the WoT vision, we focus on developing Smart Web Services (SmartWS) that encapsulate 'intelligence' by implementing autonomous decision logic in order to realize or adapt services that automatically perform tasks on behalf of the user, without requiring his/her explicit involvement. SmartWS provide remote access to resources and functionalities, by relying on standard communication protocols, and also encapsulate smartness elements, as for instance:

- Context-based adaptation – automatic adjustments based on the devices' current situation;
- Cognition – for example, learning based on available data, such as previous log files, and determining optimal settings or suggesting a particular solution option;
- Inference – for example, deducing implicit knowledge based on the available data;
- Rules – for example, formal specification of common practices or established guidelines.

Therefore, SmartWS implement autonomous decision logic in order to realize services that automatically perform tasks, such as suggesting patient diagnosis, determining an optimal traveling route, or updating the temperature settings of all heaters in a house. It is only through SmartWS that we will be able to provide the added value of interoperability, scalability and integration that is needed in order to realize the WoT.

To this end, we make the following contributions:

- We provide a definition for SmartWS;
- We motivate and introduce the key characteristics of SmartWS;
- We introduce a reference implementation framework for realizing solutions based on SmartWS;
- We describe a specific use case for implementing SmartWS in the medical domain;
- We specify a maturity model for determining the quality and usability of SmartWS.

The remainder of this paper is structured as follows: In Section 2 we present the current state of the art that provides the foundation for developing SmartWS. In Section 3 we describe the main SmartWS characteristics and how we approach their development. In section 4, we provide a reference framework for realizing SmartWS and demonstrate how it can be used to support a particular use case from the medical domain in Section 5. We introduce a maturity model for classifying SmartWS in Section 6 and conclude the paper in Section 7.

With the proposed approach and the introduced framework, SmartWS can be efficiently developed and deployed.

## 2. STATE OF THE ART

Currently, there are four lines of related work that need to be considered in the context of SmartWS. These are – (i) approaches for providing remote access to functionalities and resources over the Web (Web services, Web APIs and Microservices); (ii) approaches for data interoperability and integration on top of services, (iii) approaches for encapsulating data and functionality (e.g., inference, cognition, rules) and (iv) existing approaches aiming to support Smart Services. We discuss these lines of work in the following sections.

### 2.1 Web Services, Web APIs and Microservices

Regarding providing remote access to functionalities and resources over the Web, the past few years have been marked by a trend towards a simpler approach for developing and exposing Web service and APIs – moving away from traditional services based on SOAP [8] and WSDL [9]. Instead of relying on the rather complex WS-* specification stack, current Web APIs rely directly on the interaction primitives provided by the HTTP protocol, with data payloads transmitted directly as part of the HTTP requests and responses. If the REST architectural principles [5] are enforced on top (also referred to as RESTful services), this provides for a more coordinated and constrained communication between the application client and the server. Furthermore, Microservices [10] represent a new emerging trend towards developing Web services by realizing a number of small, highly decoupled services that focus on doing a particular small task, thus facilitating a modular approach to system-building. As a result, complex applications can be composed of a number of small, independent reusable microservices.

These developments are taken one step further by Semantic Web Services (SWS) [11], which aim to reduce the manual effort required for manipulating Web services by enhancing services with semantics. The main idea behind this research is that tasks such as the discovery, negotiation, composition and invocation can have a higher level of automation, when services are enhanced with semantic descriptions of their properties. We use the research on SWS and develop it further to provide semantic APIs. Instead of being simply an endpoint associated with a communication protocol, semantic APIs benefit from the Linked Data principles [7] and semantic technologies in order to provide access to all resources linked to a particular entity or to allow the retrieving of data based on the concepts and properties that is it linked to.

In terms of documenting the interfaces, we base the development of SmartWS on a lightweight implementation of the interfaces – by using APIs, and semantically describe these to not only capture what the service does and how it works, but also what resources can be manipulated via what inputs and outputs.

*2.2 Service Data Interoperability and Integration*

Data interoperability and integration have a long history outside the field of WoT, but the adaptation to the constraints and challenges raised by achieving such tasks in the context of dynamic, distributed and interconnected systems have received little attention [12]. For instance, the traditional Extract-Transform-Load (ETL) processes rely on data transformation pipelines to create large, centralized data warehouses, on top of which analytics can be performed. This is the approach typically employed in business intelligence [13], where the data sources are relatively static and there is a need to pull all the data in the same place. A more distributed approach is one where data interfaces are being standardized to ensure a homogeneous access to various data sources. This approach is favored in the area of Web APIs, where different data sources do not need to be pulled together, but are accessed at application level through similar interfaces [14,15]. The disadvantage here is that it puts higher effort on the client/application, as it requires to identify, access and integrate each data source separately [16]. This disadvantage can be overcome by conforming to shared means for communication and data exchange.

Another approach has been introduced more recently through the Linked Data principles [7], where links are established between resources of various data sources, pushing them in a common graph that can be manipulated and processed as if it were a common, unique data source. This facilitates data integration on a large scale and also supports the reuse of openly available datasets and vocabularies.

SmartWS are based on realizing distributed access to resources, which are integrated through the definition of links between these resources. This approach alleviates the challenge of having to integrate each data source separately, since it benefits from the links between the resources where service compositions can be made directly by defining the data that is produced by one service and consumed by the next one.

*2.3 Encapsulating Data and Functionality*

SmartWS can also benefit from the notion of Research Objects [17], which was introduced in scientific disciplines to encapsulate all the necessary information for the execution of isolated scientific experiments, ensuring the reproducibility and validation of their results and preventing decay. Research Objects have been successfully used in data-intensive disciplines like Genomics and Astrophysics and were originally conceived to extend traditional publication mechanisms [18] and take scientific communities 'beyond the pdf' by aggregating essential resources related to experimental results along with publications. This includes not only the data used but also methods applied to produce and analyze that data. We adopt this approach not to encapsulate data, as is the case with research objects, but rather to include 'smartness' elements within the service interface.

*2.4 Towards Smart Services*

The importance of developing intelligent environments for providing services and products has already been recognized as part of the Smart Service World ("Smart Service Welt") [19] vision, where services and products offered over the Internet, can be used as the basis for developing new data- and service-based business models. In this particular context "Service Platforms (Smart Services, Architecture Layer 1)" are seen as the means for implementing, configuring and composing modular value added chains. We take this vision one step further and introduce SmartWS not only as the foundation for realizing distributed solutions but also as reusable intelligent building blocks for complex applications, thus directly contributing to realizing the vision of the Smart Service World.

Furthermore, there is some specific work in the context of approaches that aim to combine Web services/Web APIs and Linked Data for realizing service invocation or composition frameworks. In particular, these approaches focus on integrating existing data services (i.e., services that provide data) exposed through Web APIs, with Linked Data principles by having services consume and produce semantic data (i.e., RDF triples). For instance, Linked Data-Fu [20] enables the development of data-driven applications that facilitate the RESTful manipulation of read/write Linked Data resources. Linked Data-Fu provides a language for declaratively specifying interactions between web resources as well as an invocation engine that performs efficiently the described interactions with the web resources.

Another approach for automatically invoking Web APIs is presented by Taheriyan et al. [21] who offer a solution that allows domain experts to create semantic models of services with the help of an interactive web-based interface. Based on samples of Web API requests and a set of vocabularies, the system invokes the service and creates a model that captures the semantics of the inputs, outputs, and their relationships. Finally, Linked Data Fragments [22] offer a new approach towards combining Web services and Linked Data by offering a solution for publishing Linked Data via a queryable API. Instead of having a solution where services consume and produce Linked Data, the authors demonstrate how existing Linked Data datasets can be exposed via a service on the Web, that allows for posting queries to single or distributed data sources.

Finally, there is already some work in the context of realizing platforms that to a certain extent implement Smart Services. Lee et al. present a Smart Service Framework (SSF) [23] that focuses on supporting context adaptation by providing a design for a centralized systems that is responsible for taking the context-relevant information from mobile and service agents, processing it based on background knowledge, such as device and user profile, and adapting the offered services to better suit the current users' needs. This approach is very important since it provides a fist step towards achieving some level of smartness. Unfortunately it focuses only on context adaptation and the authors do not discuss how individual devices are registered to the platform, how the communication is realized and how data heterogeneity is handled. Another approach is presented in [24] where the authors focus in particular on wearable devices in the context of Internet of Things. The solution is based on an enterprise service bus (ESB), which aims to support the integration of different hardware platforms into a single application and to introduce a service-oriented semantic middleware solution. Thus the ESB, in combination with semantic descriptions of the attached devices, serves as a bridge for interoperability and integration of the different environments. The approach reuses SWS technologies and centralities the communication over the ESB, however, it does not focus on adding 'smartness' elements to the used devices or their interfaces. Finally, Lee et al. [25] and Beltran et al. [26] present two domain-specific solutions – one centered around weather information and one designed specifically for aggregating social web data. Both solutions use WoT technologies, enhance them with semantics in order to support data integration and interoperability, and use SWS approaches. However, they stop short of extending the device and data interfaces with encapsulated intelligence.

In the following section we describe how we benefit from the current state of the art and take developments one step further in order to introduce Smart Web Services.

## 3. FOUNDATIONS FOR SMART WEB SERVICES

SmartWS not only enable remote data access and modification, and benefit from the Linked Data principles, but also comprise 'smartness' elements in terms of context-based adaptation (e.g., changing the weather information based on the current location), cognition (e.g., learning from past experience in order to determine optimal settings), inference (e.g., automated deduction of knowledge), and rules (e.g, formal specification of heuristics or guidelines) that implement autonomous decision logic (see Figure 1).

There are a number of definitions for smart services [23]. Usually these definitions are used to describe the offerings to the users and what they can achieve by calling the service – technical or business services. Still there is no single, generally accepted definition; especially not in the context of the technical implementation of services – as Web services or APIs. Therefore, we first give our definition of SmartWS, and subsequently specify the characteristics that SmartWS should exhibit.

***SmartWS*** *are Web APIs, conforming to standard Web technologies (HTTP, URIs), that consume and produce semantic data (RDF) and encapsulate autonomous decision logic.*

SmartWS benefit from previous research and introduce the innovative aspect of encapsulating intelligence elements directly as part of the service. As a result, SmartWS automatically adjust configurations, adapt to context or trigger events, based on the input data, thus enabling the implementation of actuators on top of the smart APIs. SmartWS are Web APIs that consume and produce Linked Data and, in addition, encapsulate rules and inference, for instance, in order to automatically deduce further knowledge that is not stored in the dataset but can be derived based on the existing facts. They implement domain or use case-specific rules that can be used to make conclusions about the outputs, based on the provided input (for example, given a temperature of 20 degrees or more, the heaters should be turned off). They can also use the relationships between concepts and instances in order to make inferences. As such, instead of only providing access to resources or existing functionalities, SmartWS are intelligent enough to deduce additional knowledge, trigger events or directly update configurations.

### 3.1 Towards SmartWS

The approach followed towards realizing the vision of SmartWS is based on three main pillars

– (i) semantic technologies, (ii) remote access to resources via Web services and (iii) 'smartness' elements in terms of, for instance, context-based, adaptation, cognition, inference and rules. This combination provides an innovative, up-to-date unexplored line of research that, as demonstrated by the use case implementation (see Section 5), shows promising results.

**(i) Semantic technologies** – As already discussed, we aim to take advantage of the Linked Data principles for publishing data, which through the use of standardized vocabularies, provides links between entities and an abundance of available datasets, enable data integration and the building of rich client applications.

(ii) **Remote access to resources** – The Linked Data principles are combined with established technologies for remotely accessing resources on the Web (e. g., URIs, HTTP and REST) thus developing interfaces (i.e., services) that consume and produce Linked Data – the service inputs and outputs are given in RDF, with formalized semantic meaning, uniquely identified resources via URIs, and links between these resources. The result is a framework for automated resources querying (service discovery), integration (service composition) and use (service invocation).

(iii) **'smartness' elements** – These semantically enabled services are enriched with logic for context adaptation, cognition, inference and rules, which are implemented directly as part of the interface. Instead of having a dedicated reasoning engine, a complete machine learning approach for optimization, or a large set of rules, the goal here is to enhance services with lightweight intelligence elements that already provide added value. There are a lot of use cases, where a little intelligence within the service goes a long way (for example, an interface that provides access to a street light that takes as input the time of the day and the light intensity outside, and based on rules within the interface automatically updates the light intensity, without having to access a centralized control system first). The result is a set of intelligent reusable semantic Web-enabled interfaces that provide access to single resources and can be used to realize decentralized distributed solutions (see Section 4).

*3.2 Characteristics of SmartWS*

In this section we derive characteristics for SmartWS, based on the aspects of 'smartness' identified in the previous section (see Figure 1). These characteristics differentiate SmartWS from traditional Web services and Semantic Web Services (SWS).

***Characteristic 1. Encapsulating 'smartness'*** The need for intelligence within the building blocks of a system has already been raised and discussed in the context of autonomic computing [27], where computing systems can manage themselves, just by receiving high-level objectives as input. We adapt this concept and take it a step further by defining the feature of encapsulating 'smartness' as having services provide human-like intelligence such as decisions or reasoning. The goal is not to have machines emulate the processes of human thinking but rather to augment services with some limited lightweight degree of intelligence regarding particular tasks.

***Characteristic 2. Adding* automation** – In general, when it comes to supporting business processes, or individual activities, it comes down to completing a number of individual tasks. Some tasks can only be performed manually, i.e., by humans, while others can be either replaced or supported by machines. When tasks can be completed directly by using SmartWS, on behalf of the user but without his/her explicit involvement, this contributes to the overall automation of the system [28], and/or supported business process. For example, the heaters in a room can automatically be turned off when the temperature is above 20 degrees.

***Characteristic 3. Adding* autonomy** – This feature relates again to reducing the human involvement. However, in order for services to be autonomous, this requires to provide elements of self-actuation, such as self-monitoring, self-diagnosis, and self-actuating. If a service already implements functions for monitoring, diagnosis, and these are actuated in autonomous manner, it directly reduces the amount of required manual intervention. Furthermore, by adding autonomy, a service would also have higher adaptability to the evolving environment and contexts of a running system.

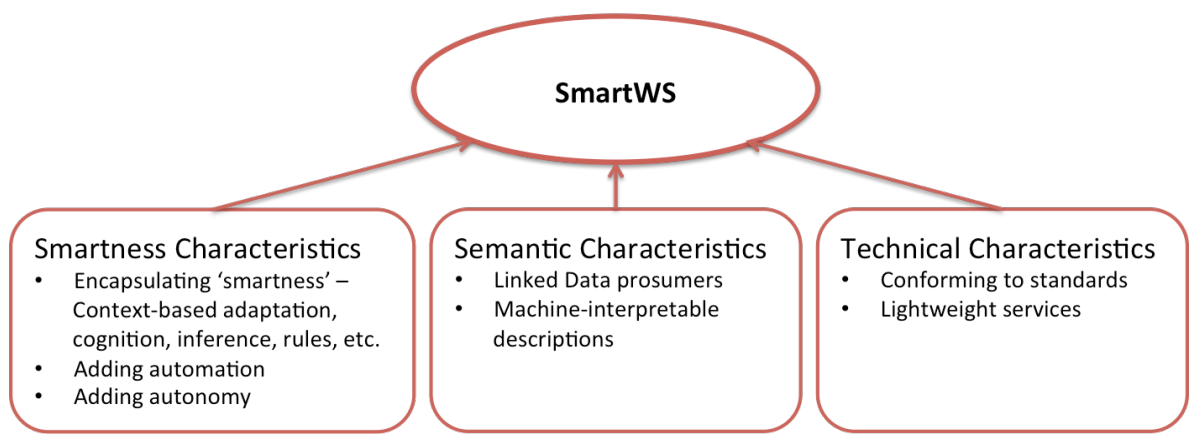

**Figure 1: Characteristics of SmartWS**

Characteristics 1-3 highlight the main differences between SmartWS, and SWS and Web services. However, SmartWS still build on semantic technologies (characteristics 4 and 5) and Web service communication standards (characteristics 6 and 7).

***Characteristic 4. Linked Data prosumers*** – SmartWS are consumers and producers of Linked Data, i.e. 'prosumers'. The inputs and outputs of the services are in RDF.

***Characteristic 5. Machine-interpretable descriptions*** – SmartWS are described in a way that enables the machine interpretation of the features of the services, including functionality, conditions for invocation, inputs, outputs and means of invocation, etc.

***Characteristic 6. Conforming to standards*** – SmartWS are realized by following standards (i.e., follow what is accepted by the general community) for communication and data exchange (HTTP, URIs, XML, JSON, RDF, etc.).

***Characteristic 7. Lightweight services*** – SmartWS are implemented as Web APIs, relying on the use of HTTP and URIs, and conforming to REST maturity level of at least 2 [29] (HTTP for message transport, XML/JSON as data format, use of resources and URIs, use of HTTP verbs).

We refer to these characteristics in more detail in order to define a maturity model for the 'smartness' of services, given in Section 6.

## 4. ARCHITECTURAL DESIGN OF A FRAMEWORK BASED ON SMART SERVICES

The development of SmartWS, based on the characteristics described in the previous section, can be assisted by providing guidelines, checklists and a reference architecture for implementing use case-based solutions. To this end, we present the design of a SmartWS framework, in terms of an architectural view. We provide a proof-of-concept implementation in the following section (see Section 5).

As it can be seen in Figure 2, we favor a basic three-tier solution. The bottom tier consist of all the data sources, hardware pieces, such as devices, wearables, sensors, algorithms, and software components, that should be made available as SmartWS. The middle layer consist of SmartWS that wrap the elements of the bottom layer behind a common interface, enhance it with semantics, and add further 'smartness'. The top layer represents the client side, which is facilitated via composite applications that combine the offered SmartWS into simple or complex processes.

The architectural approach that we follow is based on principles introduced by service oriented architectures (SOA) [30] and by integration systems [31, 32]. In the context of integration systems, instead of introducing wrappers and mediators, we use Web services as a way to wrap sources and realize the required mediation functionalities as part of the SmartWS implementation. Therefore, we support development approaches based on SOA and at the same time facilitate the benefits offered by an integration system. In the following, we describe each of the three layers in more detail.

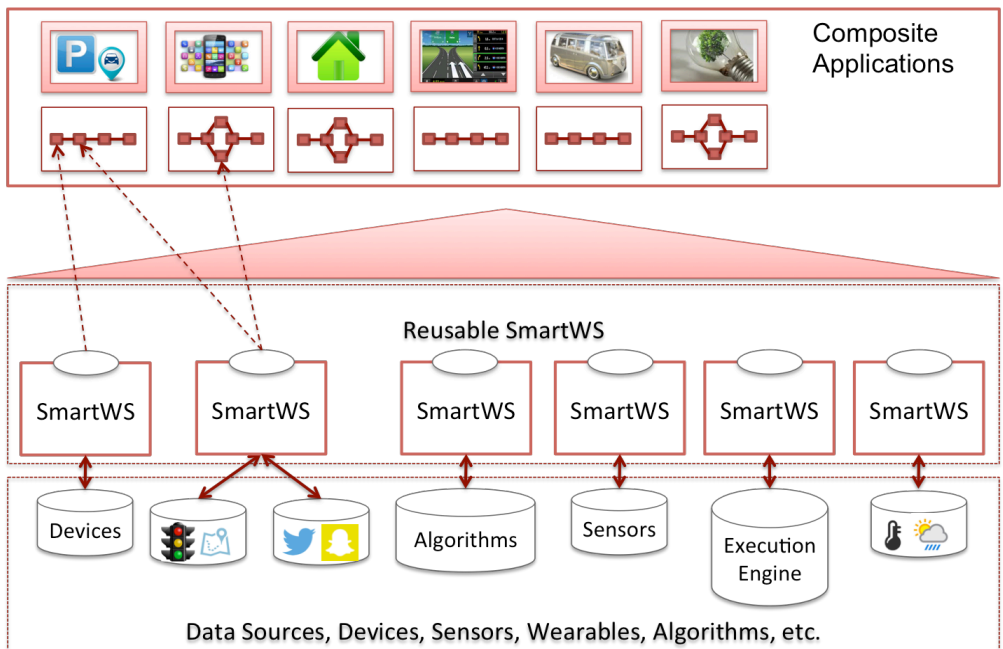

**Figure 2: Reference SmartWS Architecture**

## 4.1 Sources Layer

The sources layer consists of two main types of resources – data-centric resources and operation-centric resources. Data-centric resources are resources that are mainly related to data access and manipulation. These include, for example, repositories, databases, as well as devices and sensors that result in data streams. Therefore, we consider both static data as well as dynamic sources with a time-relevant component. Operation-centric resources focus on providing certain functionality. As examples we can have, actual Web services, algorithms, and computational software components. In addition, since any implementation can be exposed as a SmartWS in order to facilitate the development of modular distributed applications, we also advocate the approach that the actual service execution engine is made available as a SmartWS itself. The benefits are multiple, including the fact that client applications need to implement only one common interface – that of a SmartWS.

The differentiation between data-centric and operation-centric resources is not strictly required and in some instance can be difficult to determine. However, it is helpful for easing the subsequent implementation of the wrappers, in the SmartWS layer, since operation-centric sources need to be mapped to resources first in order to be able to define RESTful interfaces. While for data-centric ones this might still be required in some cases, usually the interface definition is easier.

The sources layer is extendible to comprise additional types of sources, besides the ones listed here. The only restriction is that they should be wrapable behind a SmartWS interface.

## 4.2 SmartWS Layer

The SmartWS layer realizes the implementation for facilitating all the characteristics described in Section 3. This starts with implementing wrappers for the individual sources, in order to be able to expose them via a RESTful interface. The thus realized interfaces are enhanced with semantics by creating a semantic service description and defining the inputs and outputs in terms of Linked Data concepts. Finally, 'smartness' elements are added, which can be simple rules (if input *temperature* > 20, directly set output to *off*, without further accessing the underlying resources) or more complex functions such as preprocessing the input that is sent to the source layer component, based on previously collected data.

The SmartWS can be registered in a centralized repository, as in the use case described in the next section, or can be stored in distributed repositories, for example, based on the specific provider.

## 4.3 Applications Layer

The applications layer is realized by calling individual SmartWS, creating simple compositions (sequential calling of the services) or even complex processes. The business logic that implements a workflow based on SmartWS can be encoded directly in the client application or can be controlled by another SmartWS that provides access to an execution engine. This is the layer where the actual benefit of using SmartWS becomes visible, since tasks that would normally involve some manual effort, are now completed automatically. This includes not only direct user involvement (such as, for example, turning a heater off) but also the work that needs to be invested by developers in order to implement decision logic on the client side.

In the following section we describe how we implement the SmartWS reference architecture in a specific use case in the medical domain.

## 5. USE CASE – REALIZING SMART SERVICES AS MEDICAL COGNITIVE APPS

We have already tested the practical applicability of our approach and the introduced reference architecture by exploring the possibility of realizing SmartWS in the medical domain [33, 34]. In this context, we introduced a simple form of SmartWS [35, 36], which is algorithms or applications for processing medical data, accessible via a RESTful interface and consuming and producing Linked Data. Such SmartWS can provide access to formally modeled patient data in RDF, which is exposed by publishing and interlinking individual patient records by applying Linked Data principles [37]. Furthermore, SmartWS can capture medical guidelines by describing them as formalized rules in RDF or can encapsulate processing algorithms for medical imagery. The result is a distributed Web architecture for medical diagnostic systems, which support physicians while diagnosing, based on multiple SmartWS. We see this as a use case-based poof-of-concept, where a few SmartWS were implemented to test the practical applicability.

### 5.1 Scenario – Tumor Progression Mapping

Our work is situated within the cognition-guided surgery project SFB/Transregio125[1], which is developing assistant systems for surgeons. Here, empirical knowledge, facts and patient data are being combined to identify and characterize the current medical situation and its needs, and eventually perform appropriate actions. In the context of the project, we have applied our system to the Tumor Progression Mapping use case. Tumor Progression Mapping (TPM) is an approach to visualize the timely progression of brain tumors for radiologists. The process of generating a TPM produces numerous images over time, exhibiting different characteristics. Radiologists want to see the development of the glioblastoma since the last surgery but this prerequisites tedious and complex tasks.

---
[1] http://www.cognitionguidedsurgery.de

Therefore we aim to automate the workflow for tumor progression mapping, which is as follows. First, the images are stored in a centralized imagery system, and converted into a common format. A mask for the brain region is created in the next step (Brain Mask Generation, Figure 3.), ensuring that the subsequent tasks are not influenced by bones or other structures. All prevalent brain images of a patient are then spatially registered (Batched Folder Registration). The following normalization task adapts the intensities of MRI scans, thus yielding similar values for similar tissue types (Robust or Standard Normalization). If additional annotations for a patient are available, the normalization becomes more robust. Finally, the TPM can be created (Tumor Segmentation). Optional additional steps can be automatic tumor segmentation and subsequent integration into the map (Map Generation).

directly available as RDF under persistent URIs. The SMW enables domain experts to semantically annotate their use cases, algorithms and data, and also to formally define their interaction. Another component of the knowledge base is the Semantic Patient Data Store. An instance of the XNAT[2] platform – an open source imaging informatics software platform – stores instance data for different medical departments and makes them accessible. As it can be seen, the knowledge base is not centralized, but includes individual databases (with data in its proprietary format – for example, XML, GIF, etc.) that are linked, annotated and accessible through the SMW. This provides for a lot of flexibility and extensibility.

2. **SmartWS** that provide access to medical algorithms, annotated with semantic metadata and wrapped behind a common interface, in order to be remotely accessible over HTTP.

3. An **Execution Engine,** which automatically finds, initializes and runs the algorithms, based on the information stored in the knowledge base.

The newly introduced concept of SmartWS is at the core of the system. Instead of predefining or hardcoding the sequence of the used medical

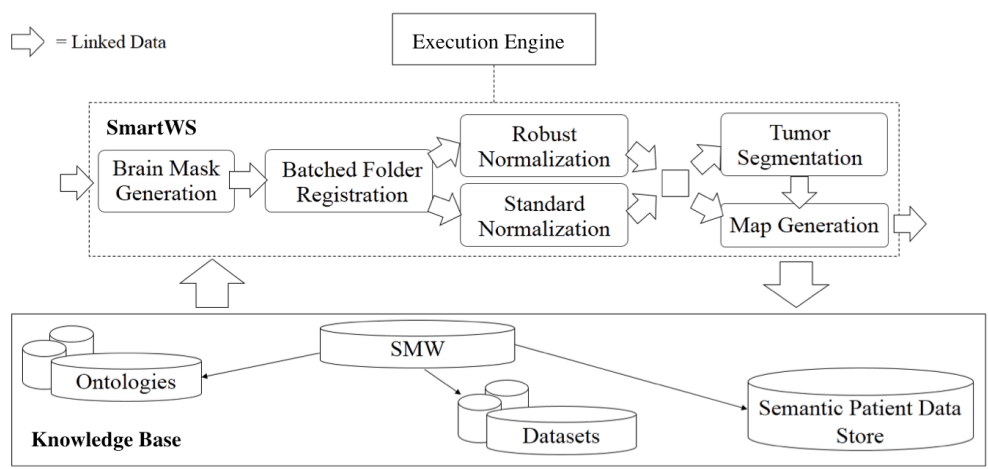

Figure 3: Brain Progression Maps Realized via SmartWS

*5.2 System Components*

In this section we describe the system design for supporting the automated TPM processing pipelines. We enable medical interpretation algorithms, such as image preprocessors, that are exposed as SmartWS to automatically run when needed in potential ad-hoc workflows. This is realized by implementing the following components (see Figure 3):

1. A **Knowledge Base** containing the medical algorithms, the data to be processed as well as training and test data for supporting individual use cases. The main entry point is a Semantic MediaWiki (SMW), which is used to enable semi-structured annotations of information from both the medical and technical worlds, which is then interpretation algorithms, we build the processing pipeline on the fly. In particular, we adopt successful Semantic Web technologies in order to create semantic annotations for algorithms, and the data they are consuming and producing. Based on these annotations we know what data is required to execute an algorithm, what this algorithm does and what are the outputs produced.

*5.3 Describing SmartWS*

A key element for supporting the automated composition of SmartWS is the formalized description of the algorithm properties. It consists of functional and non-functional properties.

[2] http://www.xnat.org

Non-functional properties comprise the following information:
- *Name*: A unique name or identifier for the algorithm within the project.
- *Contributors*: A list of contributors to the algorithm.
- *Description*: A high-level textual description of the algorithm's functionality.
- *Evaluation Metrics*: Possible evaluation metrics for the algorithm results. These are very important, since we automatically determine, which algorithm to use in the composition, based on these evaluation metrics.
- *Source Code*: Links to code repositories of the algorithm.
- *Implementation languages*: A complete list of the languages, in which the algorithm is available.
- *Service Endpoint*: This URI where the Cognitive App is executable.
- *Example Requests*: A list of URIs pointing to exemplary requests of the Cognitive App in any RDF serialization.
- *Example Responses*: A list of URIs pointing to exemplary responses of the Cognitive App in any RDF serialization.

Functional properties consist of concrete inputs and outputs of the algorithm and pre- and postconditions of the execution:
- *Inputs*: The inputs of the algorithm are either resources of type file or resources of type parameter. These resources must provide further information about their data type, about the concepts occurring in the input, about the physical format and if they are required for the execution.
- *Preconditions*: Every input must be part of a precondition to be able to specify additional constraints the algorithm has on the inputs and additional features the input should have for the algorithm to work well. Figure 4 shows the pre- and postconditions for the Brain Mask Generation algorithm, which takes as input a headscan and two reference images (a brain atlas mask and a brain atlas image). This processing step outputs the brain image and brain mask of the headscan.
- *Outputs*: The description of the outputs has the same properties as the one of the inputs.
- *Postconditions*: The description of the postconditions has the same properties as the one of the preconditions. The features depict the implications on the output, in case the algorithm was executed (see Figure 4).
- *Algorithm Class*: The type of algorithm, based on a controlled taxonomy of algorithms.

The combination of pre-, post conditions, and algorithm class enable us to automatically select suitable algorithms for completing a particular task (or task sequences). Central for our 'smartness' aspect is stipulating evaluation metrics for an algorithm. This feature depicts if and how the system can automatically quantify results based on training samples or approximate them by certain variables. As a result, algorithms that are better suited for performing a certain tasks (based on the evaluation metrics) are automatically selected and favored over others.

As a result, the information needed to run a SmartWS is directly encoded in the semantic service description. The precondition of a SmartWS specifically states what data is needed to execute the algorithm and the algorithm class enables to query for specific types of algorithms. We leverage the declarative nature of the algorithm descriptions and execute the algorithms reactively on a data-driven basis (the outputs of a SmartWS are stored back in the knowledge base and used to execute all, or a predefined subset of, SmartWS that can use these outputs as inputs; the new outputs are stored again and new SmartWS can be executed, and so on). This implies that no workflows are manually defined.

### 5.4 Realizing TPM as Data-driven Composition of SmartWS

For the execution of the SmartWS we use the Linked Data-Fu engine [20]. It uses rules to manage and define the interaction with resources on the Web (in our case the execution of SmartWS) and to virtually integrate distributed data sources. For the knowledge base, we defined Linked Data-Fu rules to search through all ontologically annotated projects, patients and files in order to select all the patient-relevant data. The composition of individual SmartWS into a TPM processing pipeline is realized as follows – the preconditions of the SmartWS are used as IF conditions for the rules. Based on the rules, the execution engine gathers all required data and then executes all SmartWS, whose preconditions are fulfilled. The results of the executed SmartWS are stored back into the knowledge base. Therefore, the knowledgebase is enriched after every execution and this enables new SmartWS to be called. The processing pipeline can be controlled, by predefining that we want to process only a single patient, a particular

```
PREFIX rdf:        <http://www.w3.org/1999/02/22-rdf-syntax-ns#>
PREFIX dc:         <http://purl.org/dc/elements/1.1/>
PREFIX sp:         <http://surgipedia.sfb 25.de/wiki/Special:URIResolver/>

?inputImage      rdf:type     sp:Category-3AHeadscan.           ?brainImage    rdf:type     sp:Category-3ABrainImage.
?inputImage      dc:format    "image/nrrd".                      ?brainImage    dc:format    "image/nrrd".
?brainAtlasImage rdf:type     sp:Category-3ABrainAtlasImage.    ?brainMask     rdf:type     sp:Category-3ABrainMask.
?brainAtlasImage dc:format    "image/mha".                       ?brainMask     dc:format    "image/nrrd".
?brainAtlasMask  rdf:type     sp:Category-3ABrainAtlasMask.
?brainAtlasMask  dc:format    "image/mha".
```

**Figure 4: Pre- and postconditions for Brain Mask Generation Algorithm**

set of images, only specific algorithms.

The implementation is also very flexible in terms of changing and extending the part of the knowledge based, which is being considered. Should further data sources be integrated, the execution engine has only to be enriched with more data-dependent rules. This also covers real-time scenarios, in which continuous data streams have to be polled.

## 6. MATURITY MODEL FOR SMARTWS

Based on the previously introduced characteristics of SmartWS and taking into account the experience gained through designing and implementing the TPM processing pipelines, we have developed a maturity model for determining the level of 'smartness' of services. The model directly relies on the SmartWS characteristics that we defined in Section 3 and captures the level of automation, autonomy and intelligent decision support that the services provide.

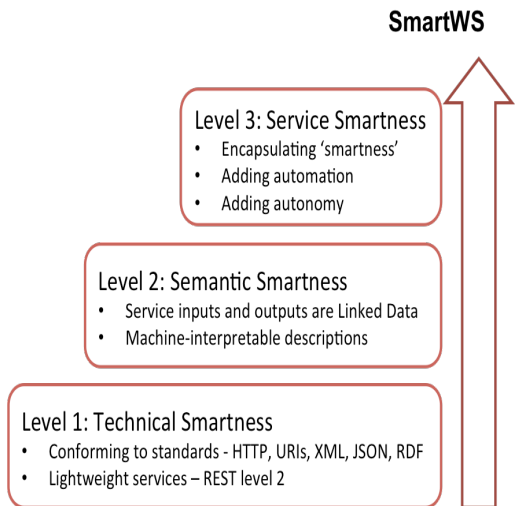

**Figure 5: Maturity Model for SmartWS**

As shown in Figure 5, we distinguish between 3 levels of 'smartness'. Level 1 is the technical level, which ensures the good accessibility and integration in terms of using of standardized access mechanisms. Level 1 of technical 'smartness' means that services are implemented as RESTful APIs, using URIs for resource identification, HTTP for communication and message transmission, as well as standard formats for data exchange, such as XML, JSON and RDF. As a result client developers know how to implement the communication with the services, solutions are more reusable and there is a direct benefit from sticking to standardized Web technologies.

Level 2 of the maturity model adds semantics in terms of both the data produced and consumed by the services as well as the actual service description. As a result services can automatically by found based on the needed functionality or the available input/expected output data. Furthermore, compositions can be made by benefiting from the Linked Data characteristics of the inputs and outputs. Similarly, already developed approaches in the context of SWS can be reused and applied. Finally, the automated execution of individual SmartWS and of compositions is supported by REST and Linked Data execution engines, such as the Lined Data-Fu engine.

Level 3 of the maturity model captures the actual added value of SmartWS. Services encapsulate 'smartness' elements (such as context adaptation, cognition, inference and reasoning), which are implemented directly as part of the service interface. Therefore, the services have their own decision logic and require less user involvement, since the SmartWS act directly without requiring further actions. For example, the heaters in a room can directly be turned off, if the temperature is above 20 degrees, without having the user do it manually. SmartWS do not only add to the level of automation but also enhance the implementation of the system. In particular, we can save on communication and data transfer between the client and the server back end, since certain outputs can be determined by the SmartWS directly based on the input (such as setting the heater to 'off') without accessing the data store or the processing component on the server. SmartWS enhance the automation and autonym of the implemented system, which are key features and a prerequisite for realizing complex use cases as envisioned by the WoT.

## 7. CONCLUSION

Current developments in the context of WoT call for new 'smarter' solutions that can handle the increasing complexity and heterogeneity resulting from the joined use of multiple mobile devices, sensors, wearables and data sources, in advanced client application scenarios. This situation is aggravated by the increasing data volumes that need to be handled and interpreted. To this end we introduce Smart Web Services (SmartWS), which encapsulate 'smartness' by implementing autonomous decision logic in order to realize or adapt services that automatically perform tasks on behalf of the users, without requiring their explicit involvement. SmartWS provide remote access to resources and functionalities, by relying on standard communication protocols, and are enhanced with semantics in terms of the inputs and outputs as well as the actual service description.

In this paper we present the key characteristics of SmartWS, and introduce a reference implementation framework for realizing systems based on SmartWS. We provide a proof-of-concept implementation of SmartWS in the medical domain and specify a maturity model for determining the quality and usability of SmartWS.

It is only through SmartWS that we will be able to provide the added value of interoperability, scalability and integration that is needed in order to realize the WoT.

ACKNOWLEDGMENT

We would like to thank Michael Goetz, Christian Weber, Marco Nolden, Klaus Maier-Hein from the German Cancer Research Center, Heidelberg, Germany for their contribution to the original development of the BTM use case published in [35, 36].